\documentclass{article}

\usepackage[final, nonatbib]{neurips_2022_ml4ps}

\usepackage[utf8]{inputenc} %
\usepackage[T1]{fontenc}    %
\usepackage{hyperref}       %
\usepackage{url}            %
\usepackage{booktabs}       %
\usepackage{amsfonts}       %
\usepackage{nicefrac}       %
\usepackage{microtype}      %
\usepackage{xcolor}         %
\usepackage{wrapfig}
\usepackage{amsmath, amsfonts}
\usepackage{bm}
\usepackage{graphicx}
\usepackage{cleveref}
\usepackage{fontawesome}

\usepackage{tikz}
\usetikzlibrary{bayesnet}
\usetikzlibrary{arrows}

\definecolor{rb4}{HTML}{27408B}

\renewcommand{\b}{\mathbf}

\title{Normalizing Flows for Hierarchical Bayesian Analysis: A Gravitational Wave Population Study}

\author{
    David Ruhe \\
    AI4Science Lab, AMLab, Anton Pannekoek Institute \\
    University of Amsterdam \\
    \texttt{d.ruhe@uva.nl} 
   \And
   Kaze W. K. Wong \\
   Center for Computational Astrophysics \\
    Flatiron Institute\\
  \texttt{kwong@flatironinstitute.org}
   \AND
   Miles Cranmer \\
   Department of Astrophysical Sciences \\
   Princeton University \\
   \texttt{mcranmer@princeton.edu}
   \And
   Patrick Forré \\
   AI4Science Lab, AMLab\\
   University of Amsterdam\\
   \texttt{p.d.forre@uva.nl}
}

\begin{document}

\maketitle

\begin{abstract}
We propose parameterizing the population distribution of the gravitational wave population modeling framework (Hierarchical Bayesian Analysis) with a normalizing flow.
We first demonstrate the merit of this method on illustrative experiments and then analyze four parameters of the latest LIGO/Virgo data release: primary mass, secondary mass, redshift, and effective spin.
Our results show that despite the small and notoriously noisy dataset, the posterior predictive distributions (assuming a prior over the parameters of the flow) of the observed gravitational wave population recover structure that agrees with robust previous phenomenological modeling results while being less susceptible to biases introduced by less flexible models.
Therefore, the method forms a promising flexible, reliable replacement for population inference distributions, even when data is highly noisy.
\href{https://github.com/DavidRuhe/hba_flows}{\faGithub}
\end{abstract}

\section{Introduction}
The number of detected gravitational wave events is rapidly increasing \cite{abbott2019gwtc, nitz20191, zackay2019highly, nitz20202, venumadhav2020new, abbott20211gwtc, abbott20212gwtc}.
Consequently, datasets of inferred properties such as mass, spin, and redshift have become feasible, the most recent being produced from GWTC-3 \cite{abbott20212gwtc}.
At this point, there is sufficient data for population-level analyses \cite{kovetz2017black, fishbach2018does, wysocki2019reconstructing, roulet2019constraints, smith2020inferring, galaudage2020gravitational, kapadia2020self, kimball2020black, roulet2020binary, tiwari2021vamana, abbott2021population, abbott2021population3}.
Many previous methods assume a simplistic phenomenological shape of the distribution such as  a power law.
However, as indicated in \cite{tiwari2021vamana}, it is crucial that the proposed model class can accurately represent the data;
otherwise, the reconstruction may miss essential population-level properties and correlations.
An imposed phenomenological shape can suffer from human-induced bias, leading to potentially incorrect conclusions.
In addition, the detection rate increases steeply with improved detector sensitivity, and it will be difficult to extract all the information from growing datasets. 
\cite{tiwari2021vamana} move away from this approach by proposing a mixture model of weighted Gaussians (and a power law) that is expected to be capable of modeling a variety of complex distributions.

In this work, we continue this direction, but use \emph{normalizing flows} \cite{tabak2010density, rezende2015variational, papamakarios2021normalizing} as our model class, as similarly done in \cite{leja2022new}.
The main advantage is that normalizing flows have successfully been able to represent complex data distributions in previous works and can be fit to such distributions in a straightforward manner.
We thus rid ourselves almost entirely of the (potentially biased) constraints put previously on the inference models.

Summarizing, we show that:
\begin{itemize}
    \item In the \emph{Hierarchical Bayesian Analysis} (HBA) framework (a Bayesian graphical model used in the gravitational wave community), a normalizing flow can, despite having access to minimal and noisy data, recover structure in the data that agrees with either ground truth or robust results from previous studies. 
    Further, it can reveal patterns that are easily missed by custom phenomenological modeling.
    \item Using a prior over the free parameters of the normalizing flow, we can get a posterior predictive distribution that generates interpretable confidence intervals.
\end{itemize}

\section{Method}
\begin{figure}
    \centering
    \includegraphics[width=0.49\linewidth]{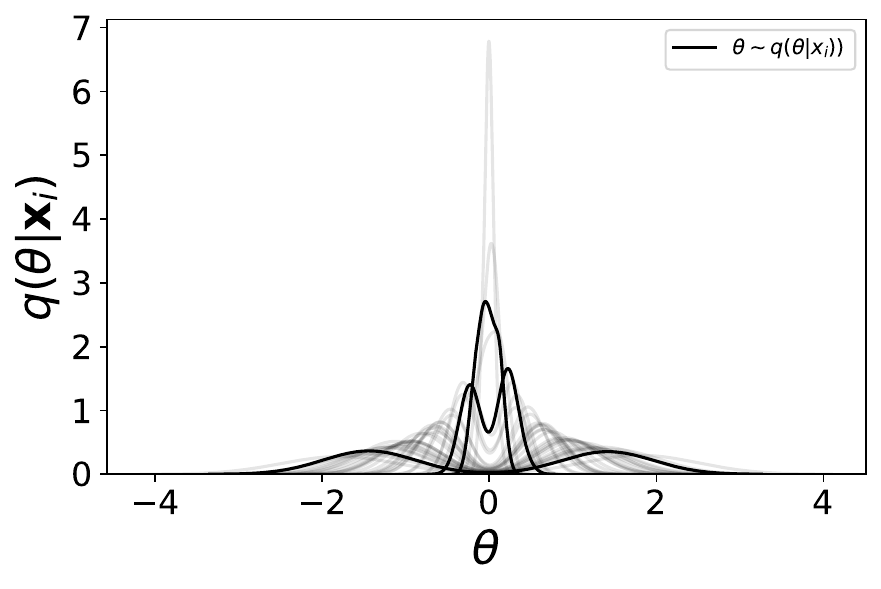}
    \includegraphics[width=0.49\linewidth]{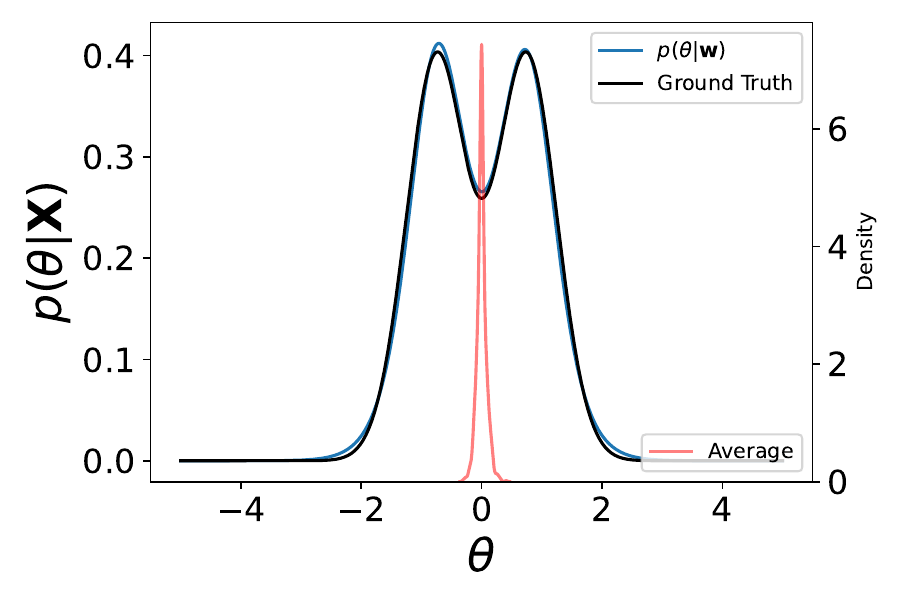}
    \caption{Left: density plots of the inferred parameters $q(\bm \theta | \b x_i)$ (some bold for visualization purposes) for each event $\b x_i$ in our simulated dataset. Right: despite the noise in the samples, the normalizing flow is able to recover the true population Gaussian mixture. We include the heuristic of averaging the noisy posterior samples, yielding an incorrect result.
    }
    \label{fig:posteriors}
\end{figure}

Let $\b x \in \mathbb{X} \subset \mathbb{R}^d$ be a gravitational wave time series and $\b X:= \{\b x_1, \dots, \b x_N \}$ be an observed dataset of such events.
We understand the physical implication of an event
$\b x$ through a vector of associated parameters $\bm \theta \in \mathbb{T}\subset \mathbb{R}^t$.
For binary mergers, these parameters include both masses, the redshift, and effective spin.
Parameter inference for individual events comes with a lot of uncertainty due to measurement noise intrinsic to the underlying time series.
Further, these posterior distributions can have a non-trivial shape.
This is exemplified in \cref{fig:posteriors} and \cref{fig:pm1}, where we show on the left the density estimates of $\bm \theta_{i}$ associated with $\b x_i$ for every $i = {1, \dots, N}$.
It is clear that heuristics such as taking the mean of the samples of the posterior distributions and fitting the population model to those is not sufficient.
Therefore, the gravitational wave community has resorted to \emph{Hierarchical Bayesian Analysis} (HBA) \cite{mandel2019extracting, gaebel2019digging, vitale2022inferring}, a Bayesian graphical model (see \cref{sec:graphical_model}) for population-level inference.
By taking into account measurement uncertainty, HBA distills the posterior samples into a single population model, which can be seen as a form of deconvolution \cite{bovy2011extreme}.
In this way, despite huge uncertainties in the parameters of individual events, we can still draw scientifically sound downstream conclusions about important characteristics of the Universe.
\subsection{Hierarchical Bayesian Analysis}
We are interested in obtaining a population model for $\bm \theta$ after observing $\b X$.
First, we introduce 
a model parameter $\b w \in \mathbb{W} \subset \mathbb{R}^w$ with a fixed prior distribution $p(\b w)$.  
For a fully specified generative model, we further need to introduce distributions $p(\bm \theta_i|\bm w)$ and $p(\b x_i|\bm \theta_i)$, $i=1,\dots,N$.
We are interested in the \emph{posterior distribution} 
$p(\b w|\b X) \propto  p(\b w) \cdot \prod_{i=1}^N \int \, d\bm \theta_i\, p(\bm \theta_i|\b w)\, p(\b x_i|\bm \theta_i)$

 and the \emph{posterior predictive distribution}
\begin{align}
    p(\bm \theta_{N + 1} | \b X)  = \int d \b w \,  p(\b w | \b X) \, p(\bm \theta_{N + 1} | \b w) = \mathbb{E}_{p(\b w | \b X)}\left[p(\bm \theta_{N+1} | \b w) \right],\label{eq:posterior_predictive}
\end{align}
which expresses the probability of observing a new event with parameters $\bm \theta_{N+1}$ given the (old) observations $\b X$ while taking into account the uncertainty in $\b w$.

While we usually do not have access to the marginal $p(\bm \theta_i)$ we can make use of a (usually physics-informed prior) distribution $q(\bm \theta_i)$, and then receive samples from the conditional $\bm \theta_{i} \sim q(\bm \theta_i|\b x_i) := \frac{p(\b x_i|\bm \theta_i)\, q(\bm \theta_i)}{q(\b x_i)} $, where $q(\b x_i):=\int\, d\bm \theta_i p(\b x_i|\bm \theta_i) \, q(\bm \theta_i)$.
I.e., we assume that we have access to samples from the individual posterior distributions.
As such, our dataset is
$
    \mathcal{D} := \left \{\left\{\bm \theta_{11}, \dots, \bm \theta_{1M_1} \right\}, \dots, \left \{\bm \theta_{N1}, \dots, \bm \theta_{NM_N} \right\} \right \},
$
where for every event $i \in (1, \dots, N)$ we have $M_i$ samples from $q(\bm \theta_i | \b x_i)$.
We are first interested in maximizing %
an approximate posterior using $\mathcal{D}$:
\begin{align}
    \b w^\star &:= \arg \max_{\b w \in \mathbb{W}} \,\tilde{p}(\b w | \b X), 
    & \tilde{p}(\b w | \b X) := \frac{p(\b w)}{Z(\b X,\mathcal{D})} \prod_{i=1}^N \left( \frac{1}{M_i} \sum_{j=1}^{M_i} \left[ \frac{p(\bm \theta_{ij} | \b w)}{q(\bm \theta_{ij})} \right]\right),
    \label{eq:objective}
\end{align}
with normalizing constant $Z(\b X,\mathcal{D})$ (derivation \cref{sec:hierarchical_bayesian_likelihood}).
A similar Monte-Carlo-based estimator is presented in \cite{vandegar2021neural}.
After obtaining $\b w^\star$ through, e.g., stochastic gradient descent, we run Markov chain Monte-Carlo around the MAP solution to obtain samples $\b w \sim \tilde{p}(\b w | \b X)$.
These samples and our model $p(\bm \theta | \b w)$ can be used to approximate the posterior predictive distribution \cref{eq:posterior_predictive}. 
It is important to model the whole posterior distribution, as the uncertainty in the parameters tells us where we can draw \emph{robust} conclusions about the underlying population.

\subsection{Normalizing Flows}
So far, we have not discussed how we parameterize $p(\bm \theta | \b w)$.
To facilitate increasingly larger datasets and to be able to model higher-dimensional distributions, the goal of this research is to explore more flexible and scalable models.
As such, we propose \emph{normalizing flows} \cite[\cref{sec:normalizing_flows}]{tabak2010density, rezende2015variational, papamakarios2021normalizing} as a suitable model class.
Normalizing flows transform a density using a chain of a smooth, invertible mappings $f: \mathbb{R}^t \to \mathbb{R}^t$ such that 
    $\bm \theta_K = f_K \circ \dots \circ f_2 \circ f_1(\bm \theta_0),
$ which constructs an arbitrarily complex density.
We parameterize each $f_k$ using a set of weights $\b w$, and construct a log-likelihood
$
    \log p(\bm \theta_K| \b w) = \log p_0(\bm \theta_0) - \sum_{k=1}^K \log \det \left \lvert \frac{\partial f_k^{\b w}}{\partial \bm \theta_k} \right \rvert,
$
which we can optimize for $\b w$ using, e.g., stochastic gradient descent.
In this work, we use \emph{planar flows} \cite{rezende2015variational} and \emph{block neural autoregressive flows} \cite{de2020block}, as we experimentally found that they are sufficiently flexible while not using too many parameters.
Furthermore, we found that the resulting distributions are \emph{smooth}, which makes them physically more plausible and allows for better automated discovery of interpretable models \cite{wong2022automated}.

\newcommand{\Xeff}{\mathcal{X}_{\text{eff}}}
\begin{figure}
    \centering
    \includegraphics[width=0.49\linewidth]{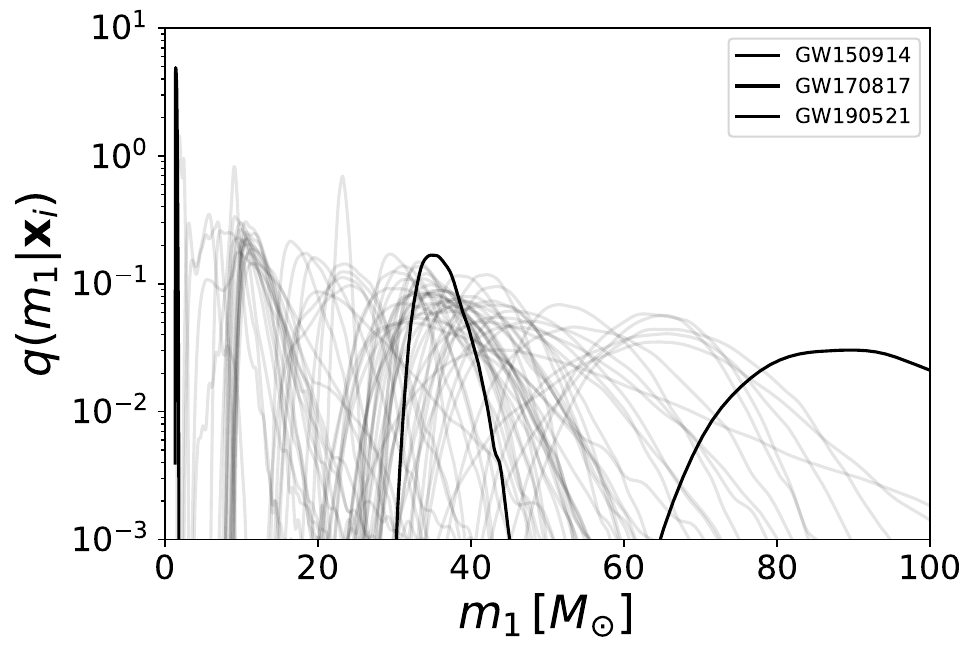}
    \includegraphics[width=0.49\linewidth]{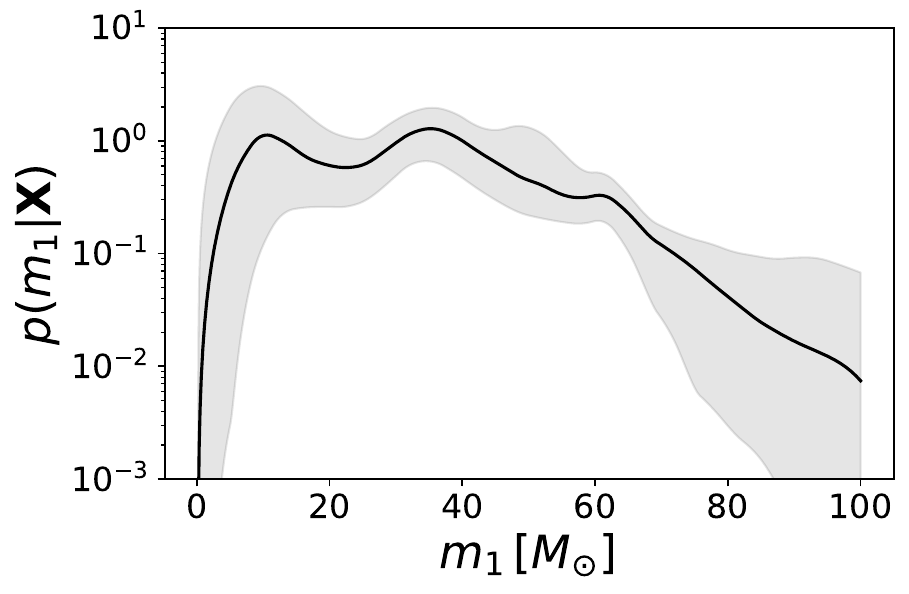}
    \caption{Left: the GWTC-3 posterior density plots. Right: our inferred population model using normalizing flows for the inspiral \emph{primary mass} $p(m_1|\b X) = \iiint dm_1\,dz\,d\Xeff\, p(\bm \theta | \b X)$ where we used a Monte Carlo estimate of \cref{eq:posterior_predictive} to generate a 90\% percentile interval.}
    \label{fig:pm1}
\end{figure}

\section{Results}
\subsection{Illustrative Examples}
In an illustrative example, we use a mixture of Gaussians as the ground truth marginal distribution over $\bm \theta$.
The observable distribution is a Gaussian with mean $\left \vert \bm \theta \right \vert$, from which we draw $\b X$. 
Using a uniform prior $q(\bm \theta)$, we obtain posterior draws $\bm \theta \sim q(\bm \theta | \b x_i)$ for all events using rejection sampling. 
Due to the noninvertible forward model, the posteriors are bimodal and noisy.
Since the true underlying population distribution and the posteriors are nontrivial, standard techniques will not be effective here.
Still, the normalizing flow can, using the Hierarchical Bayesian Analysis framework, convert the noisy posterior draws shown in \cref{fig:posteriors} (left) into a faithful population model as shown in \cref{fig:posteriors} (right).
Further, we perform a more realistic (though less clearly visualizable) experiment with synthetic data in \cref{sec:injection_experiment} where we sample data from previously suggested gravitational wave population models (e.g., by \cite{abbott2021population3}) and recover the proposed distributions.

\subsection{GWTC-3: Primary Mass Marginal}
\begin{wrapfigure}[22]{r}{0.49\textwidth}
  \begin{center}
    \vspace{-1.5cm}
    \includegraphics[width=0.48\textwidth]{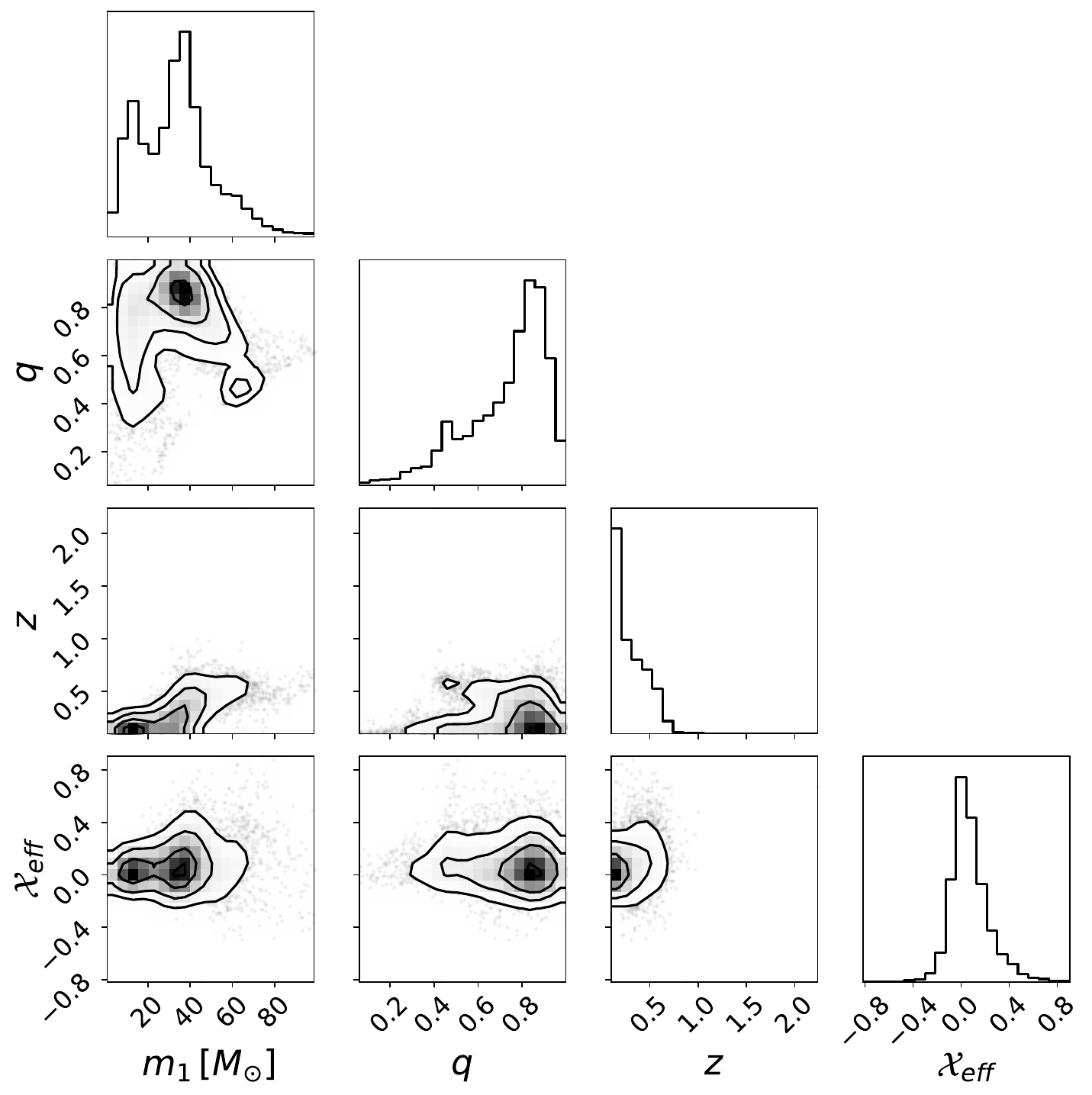}
  \end{center}
  \caption{Pair plot of joint marginals of our inferred $p(\bm \theta | \b X)$ Note that we did not take in to account selection bias effects which is especially noticeable in the redshift marginal. Hence, this is an \emph{observed population} analysis. }
    \label{fig:corner_qx}
\end{wrapfigure}
In our experiments on the GWTC-3 catalog, we consider four gravitional wave parameters: $m_1$ (\emph{primary mass}), $m_2$ (\emph{secondary mass}), $z$ (\emph{redshift}) and $\Xeff$ (\emph{effective inspiral spin}).
I.e., $\bm \theta := (m_1, m_2, z, \Xeff)$ and $t=4$.
Note that the number of recorded gravitational wave events is currently rather small ($N=76$). 
Still, we show that we can retrieve structure from the data that agrees with previous studies.
After getting $p(\bm \theta | \b w^\star)$ by maximizing \cref{eq:objective}, we obtain, using Hamiltonian Monte Carlo \cite{neal2011mcmc}, $L=10,000$ samples $\b w \sim p(\b w | \bm \theta)$, starting the chain at $\b w^*$.
Next, we obtain the posterior predictive distribution by the approximation $p(\bm \theta | \b X) \approx \mathbb{E}_{p(\b w | \b X)}[p(\bm \theta | \b w)] \approx \frac{1}{L} \sum_{l=1}^L p(\bm \theta | \b w_l)$.
We show the resulting \emph{observed population} posterior predictive marginal $p(m_1|\b X) = \iiint dm_2\,dz\,d\Xeff\, p(\bm \theta | \b X)  = \iiint dm_2\,dz\,d\Xeff \, p(m1, m2, z, \Xeff| \b X)$ in \cref{fig:pm1}.

Note, again, that we did not impose any strong restrictions on the model class; this result was fully recovered from just the data.
Overall, we see (apart from selection bias effects \cite{mandel2019extracting}) similarity with, e.g., the \textsc{Power Law + Peak} and \textsc{Multi-Peak} model of \cite{abbott2021population, abbott2021population3}.
Specifically, we observe distinct peaks around $m_1 \sim  10 \, M_{\odot}$,  $m_1 \sim  40 \, M_{\odot}$, and $m_1 \sim  60 \, M_{\odot}$. 
Finally, we also reassuringly observe that the uncertainty contracts or expands in the parts where we expect it to.
I.e., it expands at regions where we do not have many observations, and contracts in the regions where we do (e.g., around the high-mass modes).

\subsection{GWTC-3: Observed Population Distribution}

We display the marginal joint distributions in \cref{fig:corner_qx}.
Here, we define $q:= \frac{m_2}{m_1}$ as the ratio of secondary to primary mass of the binary.
Note this figure shows the \emph{observed population} results.
This is especially noticeable in the redshift marginal $z$, which is heavily biased towards low-redshift events.
Further, we expect more lower primary mass events $m_1$. 
$\Xeff$ and $q$ are expected to be less affected by not including the selection function.
To draw conclusions about the true population, one should include \emph{selection bias effects}. 
Optimization was highly unstable when including the selection function in the objective.
Hence, including selection bias effects was left for future investigation.
Also, $q$ is expected to peak at $q=1$, with an infinitely steep cutoff at $q>1$.
We did not observe this due to the smoothness of the flow.
Future work could look into reparameterizations to enable this behavior.

We can tentatively draw a few conclusions from the observed population figure.
Interestingly, it suggests that $m_1$ and $q$ do not simply correlate through a power law.
Instead, the distribution shows an inverted V shape.
The redshift to mass correlations seem plausible and consistent.
The effective spin marginal also aligns with previous literature.
In \cite{mckernan2022ligo} is was noted that $q$ does not correlate positively with $\Xeff$, which is also confirmed here.

\section{Discussion}
We parameterized the population model in the Hierarchical Bayesian Analysis framework with a normalizing flow.
In illustrative experiments, we confirmed that the model can retrieve the actual population density under nontrivial posterior samples. 
On gravitational wave data from GWTC-3, we considered primary mass, secondary mass, redshift, and effective spin and recovered an observed population model that agrees with robust previous phenomenological modeling results despite the dataset being small and highly noisy.
This paves the way for less constrained, tractable, and flexible population-level inference in noisy settings, especially once more higher-dimensional data become available, allowing for automated discovery of structure in the data.

\section{Broader Impact}
This research proposed to combine normalizing flows with the HBA framework. 
The current work found an application to gravitational wave event data, but in principle, the method can be implemented in many fields of science. While this work purely aims to aid scientific discovery, the approach can also be used in non-scientific settings dealing with noisy observations, which can pose privacy-related issues depending on the data type and source.
As such, we encourage more data privacy awareness and establishing data protection policies. 

\section{Acknowledgements}
This work was largely performed as part of the machine learning summer program of the Flatiron Institute. 
We thank Ralph Wijers for helpful feedback.
We would like to thank the scientific software development community, without whom this work would not be possible \cite{hunter2007matplotlib, van2011numpy, pedregosa2011scikit, bingham2019pyro, paszke2019pytorch, virtanen2020scipy, cobb2020scaling, waskom2021seaborn}.

\bibliographystyle{plain}
\bibliography{bib}

\appendix
\section{Graphical Model}
\label{sec:graphical_model}
\begin{figure}[h!]
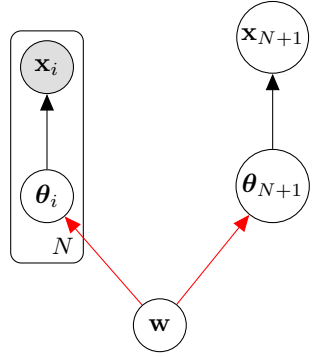

  \centering
  \tikz{
  \node[latent](w){$\b w$}; %
  \node[latent, above= of w, xshift=-1.5cm](t){$\bm \theta_i$}; %
  \node[obs, above=of t](x){$\b x_i$}; %
  \node[latent, above=of w, xshift=1.5cm](ts){$\bm \theta_{N+1}$}; %
  \node[latent, above=of ts](xs){$\b x_{N+1}$}; %
  \plate{plate}{(t)(x)}{$N$}; %
  \edge[color=red]{w}{t}; %
  \edge{t}{x}; %
  \edge[color=red]{w}{ts}; %
  \edge{ts}{xs}
     }
\caption{Graphical model of Hierarchical Bayesian Analysis. The red arrows are the ones that our normalizing flow parameterizes.}
\label{fig:gm_hba}
\end{figure}

The graphical we use is shown in \cref{fig:gm_hba}.
We colored the arrows that our normalizing flow parameterizes red.
To estimate $p(\bm \theta_{N+1} \mid \b X)$, samples $\bm \theta_i \sim q(\bm \theta_i \mid \b x_i)$ are inferred for all $\b x_i$ in the first stage.
In this work we assume we have those samples and use them to estimate the remaining densities in \cref{eq:posterior_predictive} through the objective \cref{eq:objective}.

\section{Hierarchical Bayesian Likelihood}

\label{sec:hierarchical_bayesian_likelihood}
Using $q(\bm \theta_i|\b x_i) := \frac{p(\b x_i|\bm \theta_i)\, q(\bm \theta_i)}{q(\b x_i)}$, we set
\begin{align}
    \hat{p}(\b x | \b w) &:=  \int d \bm \theta \, \left[q(\bm \theta | \b x) \cdot \frac{q(\b x)}{q(\bm \theta)}\right] \, p(\bm \theta | \b w) \\
    &=  q(\b x) \cdot \mathbb{E}_{q(\bm \theta | \b x)}  \left[ \frac{p(\bm \theta | \b w)}{q(\bm \theta)} \right].
\end{align}
Note that $q(\b x_i)$ is a constant with respect to our model parameters $\b w$.
Hence,
\begin{align}
\hat{p}(\b w | \b X) &:= \frac{p(\b w)}{Z(\b X, \mathcal{D})}\prod_{i=1}^N \mathbb{E}_{q(\bm \theta | \b x_i)}  \left[ \frac{p(\bm \theta | \b w)}{q(\bm \theta)} \right]  \\
&\approx \frac{p(\b w)}{Z(\b X,\mathcal{D})} \prod_{i=1}^N \left( \frac{1}{M_i} \sum_{j=1}^{M_i} \left[ \frac{p(\bm \theta_{j} | \b w)}{q(\bm \theta_{j})} \right]\right) =: \tilde{p}(\b w | \b X)
\end{align}
\section{Normalizing Flows}
\label{sec:normalizing_flows}
To facilitate increasingly larger datasets and to be able to model higher-dimensional distributions, the goal of this research is to explore more flexible and scalable models.
As such, we propose \emph{normalizing flows}\cite{tabak2010density, rezende2015variational, papamakarios2021normalizing} as a suitable model class.
Normalizing flows transform a density using a smooth, invertible mapping $f: \mathbb{R}^t \to \mathbb{R}^t$ with inverse $g:=f^{-1}$.
As such, if we transform a random variable $\bm \theta$, then $\bm \theta'$ has
\begin{align}
    p(\bm \theta') = p(\bm \theta) \left \lvert \det \frac{\partial f^{-1}}{\partial \bm \theta'} \right \rvert.
\end{align}
If we use a chain of these transformations, such that
\begin{align}
    \bm \theta_K = f_K \circ \dots \circ f_2 \circ f_1(\bm \theta_0),
\end{align}
we can construct an arbitrarily complex density.
We parameterize each $f_k$ using a set of weights $\b w$, and construct a log-likelihood
\begin{align}
    \log p(\bm \theta_K| \b w) = \log p_0(\bm \theta_0) - \sum_{k=1}^K \log \det \left \lvert \frac{\partial f_k^{\b w}}{\partial \bm \theta_k} \right \rvert,
\end{align}
which we can optimize for $\b w$ using, e.g., stochastic gradient descent.
The downside is a potentially computationally expensive Jacobian determinant.
As such, much work has been conducted on finding transformations that have computationally cheap Jacobian determinants \cite{dinh2016density, kingma2016improved, papamakarios2017masked, berg2018sylvester, kingma2018glow, hoogeboom2020convolution}.
In this work, we use \emph{planar flows} \cite{rezende2015variational} and \emph{block neural autoregressive flows} \cite{de2020block}, as we experimentally found that they are sufficiently flexible while not using too many parameters.
Furthermore, we found that the resulting distributions are (relative to some other parameterizations) \emph{smooth}, which makes them physically more plausible.

\paragraph{Planar Flow}
Planar flow uses the transformation
\begin{align}
f_k^{\b w}(\bm \theta) := \bm \theta + \b u_k \, h(\b v_k^\top \, \bm \theta + b_k),
\end{align}
where $(\b u_k, \b z_k, b_k) \in \b w$ are free parameters, and $h$ is the hyperbolic tangent.
These transformations contract and expand the inputs perpendicular to the plane defined $\b v_k^\top \bm \theta - b_k = 0$.
It can be shown that the Jacobian determinant of this mapping can be computed in $O(t)$ \cite{rezende2015variational}.
As such, we can use a sequence of these transformations to model $p(\bm \theta | \b w)$, our gravitational wave population model (\cref{eq:objective}).

\paragraph{Block Neural Autoregressive Flow}
Neural autoregressive flows decomposes a joint distribution over $\bm \theta \in \mathbb{R}^t$ into $t$ conditional distributions.
the transformations $f_k^{\b w}$ then yield lower triangular Jacobians and hence cheaply computable determinants.
Block neural autoregressive flows then parameterize $f_k^{\b w}$ using a neural network that uses block lower-triangular matrices.
Their diagonal elements are strictly positive, ensuring monotonicity.
The off-diagonal elements do not require monotonicity as they only play a role in the conditioning part of the conditional distributions.

\section{Experimental Details}
All experiments were conducted locally.
\subsection{Illustrative Example}
The ground-truth population is defined by a mixture $\bm \theta \sim \mathcal{N}(-0.75, 0.5^2)\mathcal{N}(0.75, 0.5^2)$.
The forward model is given by $\b x|\bm \theta \sim N(|\bm \theta|, 0.1^2)$.
This noninvertible forward mapping results in noisy, bimodal posteriors.
We sample $N=1024$ events. 
Using a uniform prior $q(\bm \theta) := U(-5, 5)$ and the true forward model, we draw $M_i=128$ posterior samples using rejection sampling from $q(\bm \theta | \b x_i)$ for every $\b x_i$, $i \in 1, \dots, 1024$.
We parameterize $p(\bm \theta | \b w)$ as a 4-layer planar flow and a standard Gaussian base distribution.
We use the Adam optimizer \cite{kingma2014adam} with learning rate $0.01$ so obtain $\b w^\star$.
We found empirically that optimization runs better when mini-batches are taken for both the events and also the posterior draws.
In this experiment, we batch sizes of $N':=1024$ events and $M_i':=4$.

\subsection{Synthetic Data Experiment}
\label{sec:injection_experiment}
\begin{figure}
    \centering
    \includegraphics[width=0.49\linewidth]{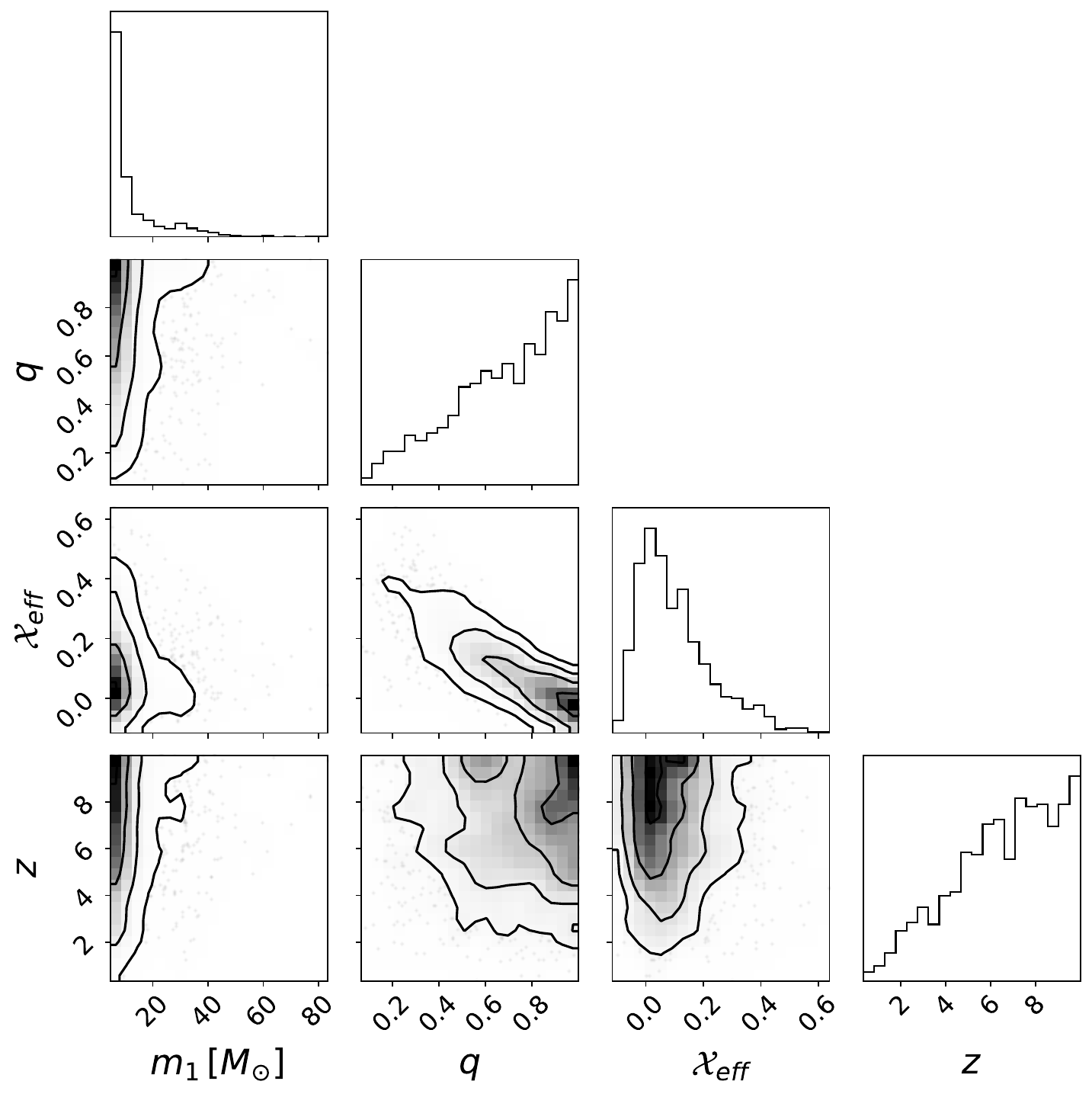}
    \includegraphics[width=0.49\linewidth]{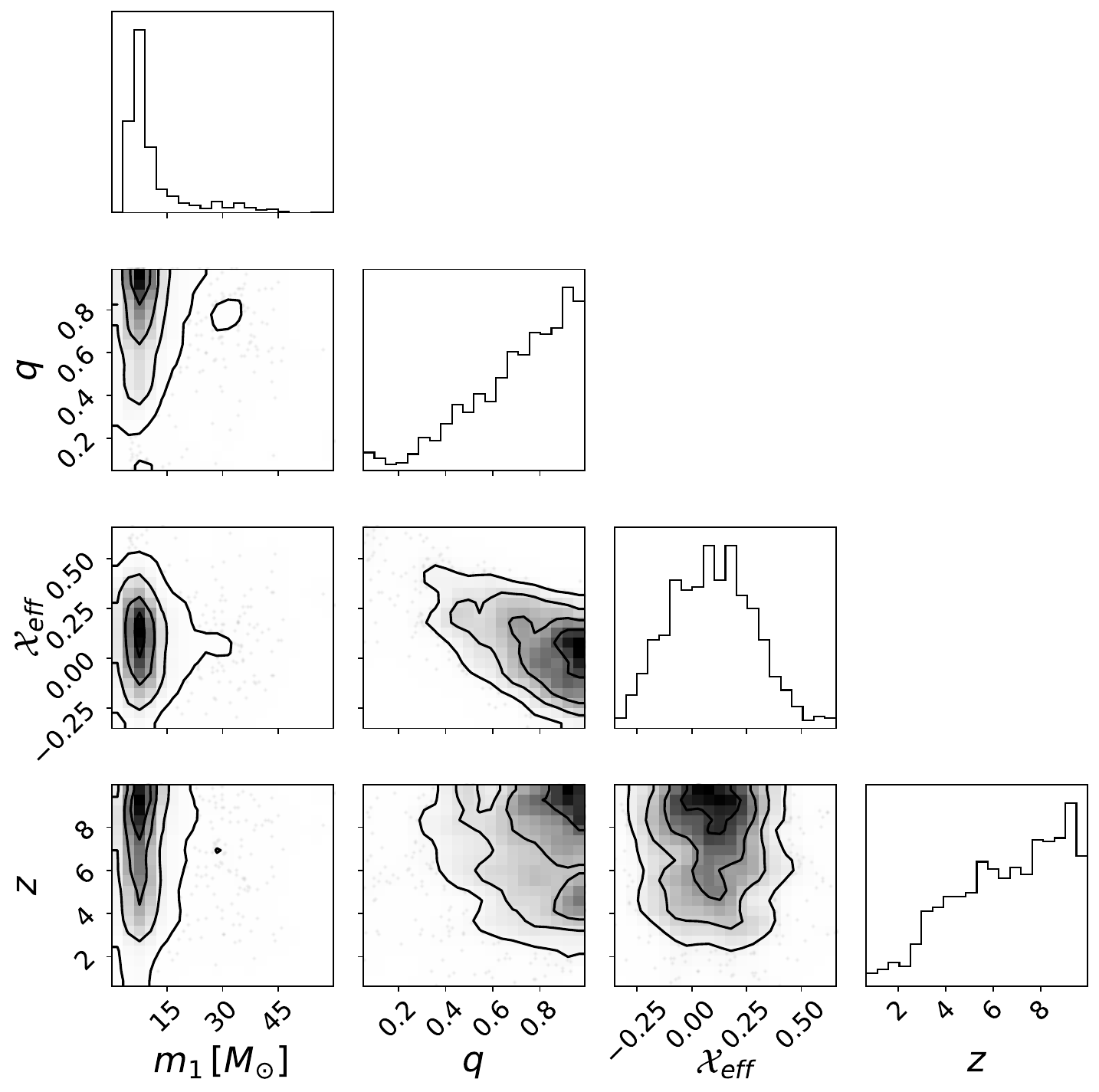}
    \caption{Left: simulated gravitational wave data. Right: recovered distributions by the normalizing flow.}
    \label{fig:injection_experiment}
\end{figure}%
Here, we simulate a population model using previously published models and investigate whether our flow-based approach can faithfully recover those distributions.
We start with the \textsc{Power Law + Peak} model for $p(m_1, m_2)$ from \cite{abbott2021population3} using the means of the posterior distributions for ground-truth population parameters.
For a conditional effective spin distribution we use \cite{callister2021ordered}, again with the mean of the inferred distributions as ground-truth parameter values.
Finally, we model the distribution of redshift using a power law with spectral index $\kappa=1.7$ following \cite{Fishbach:2018edt}.
Note that there is an extra factor in the merger coming from the comoving volume factor.
The resulting samples are shown in \cref{fig:injection_experiment} (left).
We simulate Gaussian posteriors in our Hierarchical Bayesian framework.
The distribution that the normalizing flow recovers is shown on the right in \cref{fig:injection_experiment}.
We see that the flow recovers most of the correlations and shapes of the joint distribution faithfully.

We reparameterize the $m_1$ samples to log-space and $q$ samples to logit-space.
We sample $N=1024$ events and use 32 hierarchical samples per batch. 
We used the Adam optimizer \cite{kingma2014adam} with learning rate of 0.001 and 1 layer of block neural autoregressive flow with 2 8-dimensional hidden layers. 

\subsection{GWTC-3}
We have $N=76$ gravitational wave events available.
The number of posterior samples per event ranges from $M_i=3194$ to $M_i=268806$. 
We use 10 layers of $\emph{planar flow}$, resulting in a total of $w=90$ free parameters, limiting the risk of severe over-fitting resulting in unnatural population distribution shapes.
We use mini-batches of $N':=76$ events and $M_i':=32$ posterior samples per event.
We use the Adam optimizer \cite{kingma2014adam} with learning rate 0.001. to find $\b w^\star$.

\section{Future Work}
In this study, we did not include \emph{selection bias} effects. 
Future work should adjust for this to obtain more realistic models of population parameters of gravitational waves.
We also expect that we did not fully explore all parameterizations (and hyperparameters) that recover the population model as faithfully as possible, both in terms of the normalizing flow and how the input data should be transformed for the flow to optimally learn the densities.

Further, \cite{wong2022automated} have presented a promising approach to distill black-box models into symbolic equations.
Such an approach can also be applied to our normalizing flows, yielding more interpretable versions of the presented population densities.

Finally, we used Hamiltonian Monte-Carlo to get samples $\b w \sim p(\b w | \bm \theta)$. 
This is not scalable when the number of weights increases in order to be able to fit more complex data.
To resolve this, follow-up works could explore approaches based on the \emph{Bayesian Neural Network} literature, e.g., \cite{blundell2015weight, gal2016dropout, lakshminarayanan2017simple}.

\end{document}